\documentclass[9pt,executive]{article}
\usepackage[left=1in,top=1in,right=1in,head=1in,foot=1in]{geometry}
\usepackage{times}
\usepackage{epsfig}
\usepackage{latexsym}
\usepackage{amssymb}
\usepackage{alltt}
\usepackage{color}
\usepackage{colortbl}
\definecolor{mygrey}{rgb}{0.82,0.82,0.82}

\def\presp{PRES\textsuperscript{+}}
\def\fsmd{FSMD}
\def\fsmdb{(FSMD)}
\def\algofunction{constructSetOfTransitions}
\def\presp{PRES\textsuperscript{+}}
\def\fsmd{FSMD}
\def\fsmdb{(FSMD)}
\def\algofunction{constructSetOfTransitions}
\begin{document}                                                      
\begin{center}
{\bf \LARGE  Equivalence Checking in Embedded Systems Design Verification}
\end{center}
\vspace*{24pt}
\begin{center}
 {\Large \bf Soumyadip Bandyopadhyay}\\
\vspace*{12pt} 
{\Large \bf soumyadip@cse.iitkgp.ernet.in}
\end{center}
\vspace*{24pt}

\section{Introduction}
In this paper we focus on some aspects related to modeling and formal verification of embedded systems.
Many models have been proposed to represent embedded systems \cite{sedwads} \cite{peles}. 
These models encompass a broad range of styles, characteristics, and application domains and include 
the extensions of finite state machines, data flow graphs, communication processes and Petri nets. 
In this report, we have used a {\presp} model (Petri net based Representation for Embedded Systems)
as an extension of classical Petri net model that captures concurrency, timing behaviour of
embedded systems; it allows systems to be representative in different levels of
abstraction and improves expressiveness by allowing the token to carry information
\cite{zebo}. This modeling formalism has a well defined semantics so that it supports
a precise representation of system. As a first step, we have taken an untimed {\presp}
model which captures all the features of {\presp} model except the time behaviour which have 
reported in earlier report.

   A typical synthesis flow of complex systems like VLSI circuits or embedded systems
comprises several phases. Each phase transforms/refines the input behavioural
specification (of the systems to be designed) with a view to optimize time and
physical resources. Behavioural verification involves demonstrating the equivalence
between the input behaviour and the final design which is the output of the last phase.
In computational terms, it is required to show that all the computations represented
by the input behavioural description, and exactly those, are captured by the output
description.

  Modeling using {\presp}, as discussed above, may be convenient for specifying the input
behaviour because it supports concurrency. However, there is no equivalence checking
method reported in the literature for {\presp} models to the best of our knowledge.
In contrast, equivalence checking of {\fsmd} models exist \cite{ChandanKarfa}. Although
Transformation procedure from non-pipelined version {\presp} to pipelined version {\presp} is reported \cite{zebo}. 
As a first step, we seek to hand execute our reported algorithm on a real life example and we have to translate 
two versions of {\presp} models to {\fsmd} models. 

  The rest of the paper is organized as follows. 
 Section \ref{description} presents the definition of {\presp} and {\fsmd} models.
 Section \ref{algorithm} presents Proposed algorithm for conversion from an untimed {\presp} models to an {\fsmd} models. 
 Section \ref{real life example}  presents notion of equivalence, abstraction. In this section we have also presented the
 working principal of an example of real life embedded systems.  
 Section \ref{results} verify the equivalence between initial and transformed behaviour using {\fsmd} equivalence checking method. 
 Finally, some future works are identified in Section \ref{future-work}
\section{Brief description of {\presp} and {\fsmd} model}\label{description}
Before the conversion mechanism we discuss the design representation of {\presp} models.
\subsection{ Description of {\presp} models}
  A {\presp} model is a seven tuple $N = (P, V_P, K, T, I_P, O, M_0)$,
where the members are defined as follows. The set $P = \{p_1 , p_2 , ...., p_m\}$
is a finite non-empty set of places; $V_P$: the set of variables. A place $p$ is associated with a variable $v_p$; 
therefore, $V_P = \{ v_p \mid p \in P\}$. Every place is capable of holding a token having a value. 
A token value may be of any type, such as, Boolean, integer, etc., or a user-defined type of any 
complexity (for instance, a structure, a set, or a record). The set $K$ denotes the set of all possible token types. 
Thus, $K$ is a set of sets. The set $T = \{t_1, t_2 , ...., t_n \}$  is a finite non-empty set of transitions; 
$I_P \subseteq P \times T$ is a finite non-empty set of input arcs which define the flow relation from places to 
transitions $-$ ``input'' with respect to transitions; $O \subseteq T \times P$ is a finite non empty set of output arcs 
which define the flow relation from transitions to places. A marking $M$  is the assignment of tokens to places of the net; 
hence, $M \subseteq P$. The marking of a place $p\in P$, denoted $M(p) $, is either  $0$ or $1$. For a particular marking M, 
a place p is said to be marked iff $M(p) = 1$. $M_0$ is the initial marking of the net, depicting the places having tokens initially. 
  
  The type function $\tau$: $P \rightarrow K$ associates every place $p \in P$
with a token type. 

  The pre-set  $^\circ{t}$ of a transition $t \in T$ is the set of input  places of $t$.
Thus, $^\circ{t} = \{p \in P \mid (p,t) \in I_P \}$. Similarly, the post-set  $t^\circ$
of a transition $t \in T$ is the set of output  places of $t$. So, $t^\circ = \{ p
\in P \mid (t,p) \in O \}$ and $ \forall t \in T, \forall p_1, p_2 \in t^\circ,
\tau(p_1) =  \tau(p_2)$ and $ v_{p_1} = v_{p_2}$. The subset $ V_{^\circ{t}} = \{ v_p
\mid p \in {^\circ}{t}\}$ is the set of variables associated with places from which input
arcs lead to the transition $t$. Similarly, the pre-set $^\circ{p}$  and the post-set $p
^\circ$   of a place $p \in P$ are given by $^\circ{p} = \{ t \in T \mid (t, p) \in O
\}$ and $p ^\circ = \{t \in T \mid (p, t) \in I_p\}$, respectively. 

  For every transition $t \in T$, there exists a transition  function $f_t$ associated
with $t$; that is, for all $t \in T$, $f_t$: $\tau(p_1) \times \tau(p_2) \times ....
\times \tau(p_a) \rightarrow \tau (q)$, where $^\circ{t} = \{p_1, p_2,....., p_a\}$ and
$q \in t^\circ$. The functions $f_t$'s are used to capture the functional transforms that take place of the variable associated with the output places of the transitions i.e, $v_q \Leftarrow f_t(v_{p_1}, v_{p_2}, ...v_{p_a})$.

  A transition $t \in T$ may have a guard $g_t$ associated with it. The guard of a
transition $t$ is a predicate $g_t$: $\tau(p_1) \times \tau(p_2) \times .... \times
\tau(p_a)  \rightarrow  \{0, 1\}$, where $^\circ{t} = \{p_1, p_2,..., p_a\}$ over the
variable set $V_{^\circ{t}}$. 
\subsection { Description of {\fsmd} model}
 A finite state machine with data path {\fsmdb} is a universal specification model. An
{\fsmd} is defined as an ordered tuple $F = (Q, q_0, I_F, V_F, O, f, h)$ where 

 $Q = \{q_0, q_1, ...., q_n\} $ is a finite set of control states.
 $ q_0 \in Q $ is the reset state.
 $I_F$ is the set of primary input signals.
 $V_F$ is the set of storage variables.
 $O_F$ is the set of primary output signals, $O_F  \subseteq  V_F$.
 $f$: $Q \times 2^S \rightarrow Q$ is the state transition function.
 $h$: $Q \times 2^S \rightarrow U$ is the update function of the output and the storage variables, where S and U are as defined below
 $S = \{L \cup E_R \mid L$ is the set of boolean literals of the form $b$ or $~b$, $b \in B \subseteq V$ is a boolean variable and $E_R = \{e R 0 \mid  e \in E_A \}\}$; its represent the set of status expression over $I_F \cup V$, where $E_A$ represents a set of arithmetic expression over $I_F \cup U$ of input and storage variables and $R$ is any arithmetic relation. $R \in \{=, \neq, >, \geq, <, \leq \}$.
 $U = \{x \Leftarrow e \mid$ $ x \in O_F \cup V_F$ and $e \in E_A \cup E_R \}$ represent set of storage or output assignment.

\section{Proposed algorithm for conversion from an untimed {\presp} models to an {\fsmd} models}\label{algorithm}
 Let the input {\presp} model be $N$ and the generated {\fsmd} model be $F$. For simplicity, we assume that all tokens are of integer type. i.e $\tau (p)$ = $Z$ for all $p \in P$.

 The first step of our algorithm computes the following entities in the {\fsmd} model:  $q_0, I_F, V_F, O_F, U$ and $S$.
 The algorithm then goes on to compute $Q$: the set of states; $f$: the state transition function and $h$: the update function. 
Symbolic simulation of the {\presp} model is used to compute these entities starting from the initial marking $M_0 = q_0$.
\begin{itemize}
 
 \item 
 At each step of the simulation, starting from a present marking $M (= q) \subseteq
P$ the algorithm enumerates all the possible sets of transitions of $N$ from $M$; for
each of these sets of possible transitions, it constructs the next state $(q^+)$ of $F$
from the new marking $M^+$ of the {\presp} model $N$.
 \item 
 Obtain the transition from $q$ to $q^+$ in $F$ .     
 \item
 \begin{figure}[htbp]
\centerline{\epsfig{figure=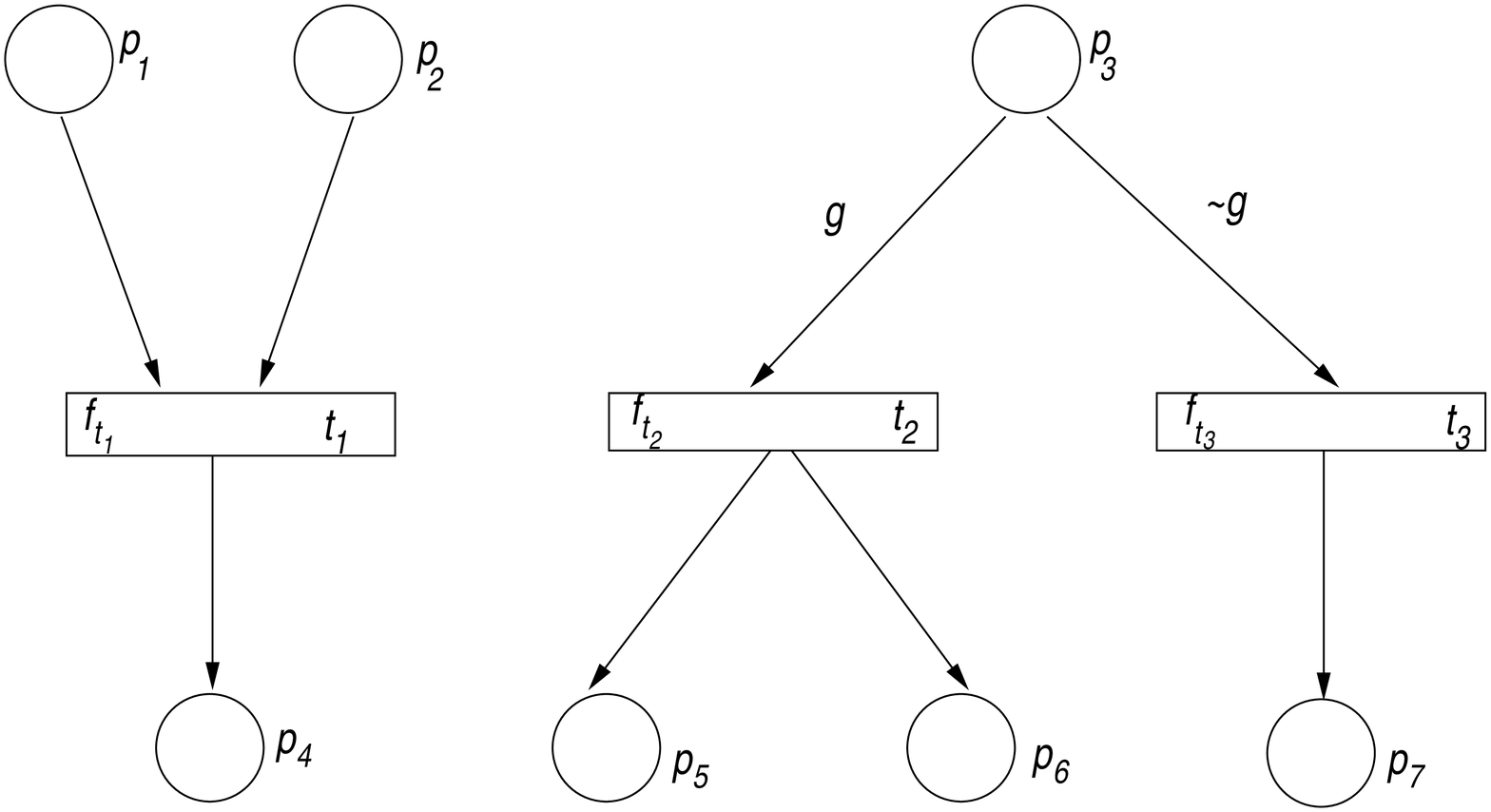,height=50mm}}
\caption{Places and transitions in a {\presp} model}
\label{fig:guards}
\end{figure}
For example, consider the scenario given in Figure \ref{fig:guards}. Let $M = \{p_1, p_2, p_3\} = q$; so the set $T_q$ of all transitions emanating from the places in M is given by $T_q = \{t_1, t_2, t_3\}$. The possible sets of transitions are $\{t_1, t_2\}$ leading to the marking $M_1^{+} = \{p_4, p_5, p_6\} = q_1^{+}$ and $\{t_1, t_3\}$ leading to the marking $M_2^{+} = \{p_4, p_7\} = q_2^{+}$. The {\fsmd} transition $(q \rightarrow q_1^{+})$ is associated with the guard condition $g$ and the {\fsmd} transition $(q \rightarrow q_2^{+})$ is associated with the guard condition $\neg g$, i.e, $f(q, g) = q_1^{+}$ and $f(q, \neg g) = q_2^{+}$. $h(q, g) : v_{p_4} \Leftarrow f_{t_1} (v_{p_1}, v_{p_2})$ and $v_{p_6} = v_{p_5} \Leftarrow f_{t_2}(v_{p_3})$. $h(q, \neg g) : v_{p_4} \Leftarrow f_{t_1} (v_{p_1}, v_{p_2})$ and $v_{p_7} \Leftarrow f_{t_3}(v_{p_7})$.



 
 \end{itemize}

{\bf Algorithm}

\noindent
{\it Steps:}\\
{\bf Step 1:} Given {\presp} model\\
\hspace*{0.42in} $q_0 \Leftarrow M_0$;\\
\hspace*{0.42in} $I_F \Leftarrow$ \{ Variables associated with $p \mid p \in M_0(p)\}$;\\
\hspace*{0.42in} $V_F \Leftarrow$ \{Variables associated with $p \mid p \notin\ M_0(p)\}$;\\ 
\hspace*{0.42in} // $O_F$ is the set of variables  associated with places from which no arcs are input 
\hspace*{0.42in} // to any transition. Therefore \\
\hspace*{0.42in} $O_F \Leftarrow$ \{Variable associated with $p \mid p ^\circ = \phi\}$;\\
\hspace*{0.42in} // $U$ is obtain from transition function of {\presp} model and variable associated
\hspace*{0.42in} // with post set of that transition. Therefore,\\ 
\hspace*{0.42in} $U \Leftarrow \{x \Leftarrow f_t^{n} (v_1, v_2, ...., v_n) \mid t \in T, f_t^{n}$ is the function associated with  
\hspace*{0.48in} $t$, $ x = v_{t ^\circ} $ and $ v_i \in v_{^\circ{t}}, 1 \leq i \leq n \}$;\\ 
\hspace*{0.42in} // $S$ is obtained from the guard conditions  of the {\presp} models. Therefore,\\
\hspace*{0.42in} $S  \Leftarrow  \{ g_t \mid t \in T \}$;\\ 
{\bf Step 2:} $Q \Leftarrow  \{q_0\}$; $Q_{new} \Leftarrow Q$; $Q_{new}^+ \Leftarrow \emptyset$;\\
{\bf Step 3:} $\forall q \in Q_{new}$ \\
\newpage
\hspace*{0.25in}{\bf Step 3.1:}$Q_{new} \Leftarrow Q_{new}$ $-$ $\{q\}$; $T_q \Leftarrow \{t \mid $  $^\circ{t}$ $\in q\}$;\\
\hspace*{0.81in}$\tau_q \Leftarrow$ {\algofunction} $(T_q)$; // $\tau_q \in 2^{T_q}$, the set of possible\\
\hspace*{2.84in} // transitions.\\
\hspace*{0.81in}$Q_{new}^q = \emptyset$, empty set, //$Q_{new}^q$: the set of next states generated \\
\hspace*{2.01in} // depending on $q$ mutually exclusive\\
\hspace*{2.01in} // depending on guard condition\\ 
\hspace*{2.01in} // associated with member of $\tau _q$.\\  
\hspace*{0.25in}{\bf Step 3.2:} $\forall T \in \tau _q$\\ 
\hspace*{0.35in} {\bf Step 3.2.1:} $q_T^{+} \Leftarrow \{t \mid t_i \in T\}$; $Q_{new} \Leftarrow  Q_{new}^q \cup \{q_T^{+}\}$;\\             
\hspace*{0.35in} {\bf Step 3.2.2:} Let $G_T$ be the set of guards associated with $t \in T$. In the table\\
\hspace*{1.04in} of the function $f$, insert entry\\
\hspace*{1.04in} $f(q, G_T) = q^+$\\
\hspace*{0.35in}{\bf Step 3.2.3:} Let $A_T$ be the set of assignments of the form\\
\hspace*{1.06in}$\{v \Leftarrow f_t(v_1, v_2, ..., v_n) \mid t \in T, \{v\} = t ^\circ, \{v_1, v_2, ..., v_n\} = ^\circ{t}$\\
\hspace*{2.36in} and $f_t$ is the function associated with $t$ \}; \\
\hspace*{1.06in} In the table of the function $h$, insert the entry $h(q, G_t) = A_T$;\\ 
\hspace*{1.06in} // members of $A_T$ are carried out in parallel\\ 
\hspace*{0.35in}{\bf Step 3.2.4:} $Q_{new}^+ \Leftarrow Q_{new}^+ \cup  Q_{new}^q$; \\        
{\bf Step 4:} // Any new state generated \\
\hspace*{0.42in} $Q_{new}^+ \Leftarrow Q_{new}^+$ $-$ $Q$;\\
\hspace*{0.42in} if $Q_{new}^+ = \emptyset$  exit;\\
\hspace*{0.48in} else \{ $Q \Leftarrow Q \cup Q_{new}^+$; $Q_{new} \Leftarrow Q_{new}^+$; $Q_{new}^+ \Leftarrow \emptyset$;\\
\hspace*{0.83in} goto {\it Step 3}\\
\hspace*{0.75in} \}\\

\begin{figure}[htbp]
\centerline{\epsfig{figure=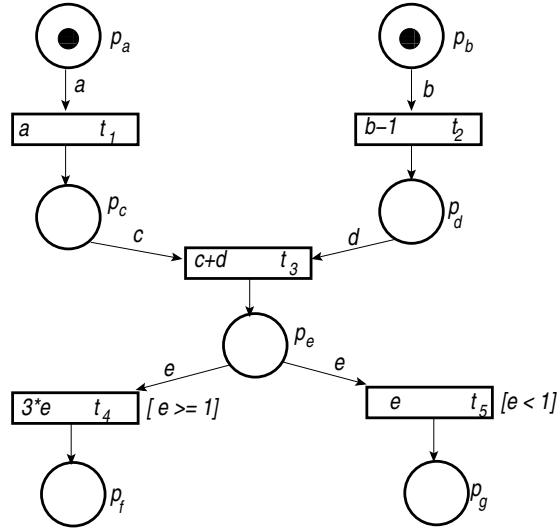,height=70mm}}
\caption{{\presp} model to be converted into {\fsmd} model}
\end{figure}
            
\begin{figure}[htbp]
\centerline{\epsfig{figure=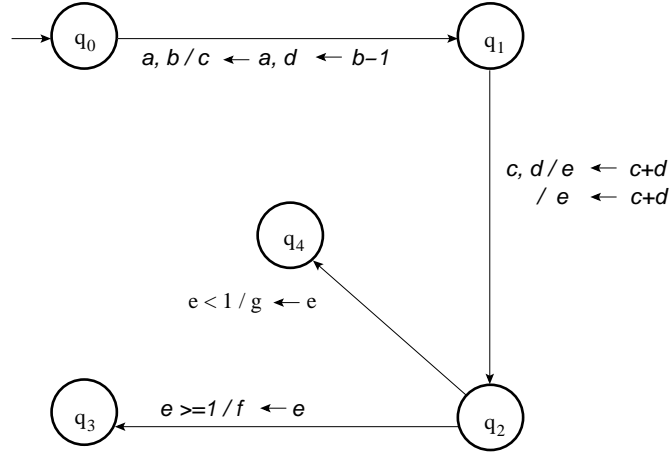,height=60mm}}
\caption{{\fsmd} model equivalent to the {\presp} model of Figure 1}

\end{figure}
 
\section{Notion of equivalence and Real life example} \label{real life example}

\subsection{Notion of equivalence between two {\presp} models}

\begin{figure}[htbp]
\centerline{\epsfig{figure=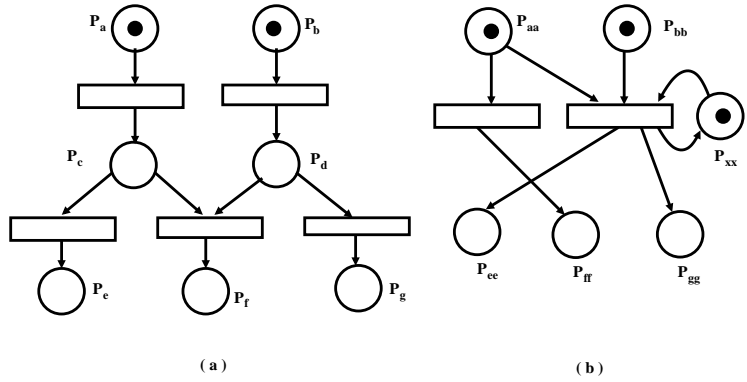,height=50mm}}
\caption{Cardinality equivalence nets}
\label{fig:n}
\end{figure}

In the synthesis process there are a  number of refinement phase. System model is transformed in each phases. 
So the validity of this transformation  depends on the equivalence between the input behaviour and the output behaviour of each phase. 
Literature \cite{zebo} has propounded three notion of equivalence - cardinality equivalence, functional equivalence, and time equivalence;
the two {\presp} models are totally equivalence iff they satisfies all these equivalence. We are dealing with untimed {\presp} hence, 
there is no need to show time equivalence.
Two {\presp} models  $N_1$ and $N_2$ are cardinality equivalence iff:
 \begin{enumerate}
   \item There exist a  one to one correspondence between the in-ports and the out-ports of $N_1$ and $N_2$ 
      i.e $f_{in}$: $inP_1$ $\leftrightarrow$ $inP_2$ and  $f_{out}$: $outP_1$ $\leftrightarrow$ $outP_2$.
   \item The Initial markings $M_{1, 0}$ and $M_{2, 0}$ of $N_1$ and $N_2$ are the same.
   \item After execution of $N_1$ and $N_2$ if the tokens are accumulated at out-ports of the each nets, there is 
         a one to one correspondence of marking at their out-ports. 
 \end{enumerate}
 For example in  Figure \ref{fig:n} in$P_1$ = \{$P_a$, $P_b$\}, out$P_1$ = \{$P_e$, $P_f$, $P_g$\}, 
in$P_2$ = \{$P_{aa}$, $P_{bb}$\} out$P_2$ = \{$P_{ee}$, $P_{ff}$, $P_{gg}$\} and $f_{in}$ and $f_{out}$ are defined by
$f_{in}$($P_a$) = $P_{aa}$, $f_{in}$($P_b$) = $P_{bb}$, $f_{out}$($P_e$) = $P_{ee}$, $f_{in}$($P_f$) = $P_{ff}$ and $f_{in}$($P_g$) = $P_{gg}$.
Second condition also satisfies the two nets. $N_1$ and $N_2$ also satisfies third condition i.e after execution of $N_1$ and $N_2$ 
all out-ports of $N_1$ and $N_2$ contains token and they are one to one correspondence. Hence two {\presp} $N_1$ and $N_2$ are cardinality 
equivalence. 
 
 Two nets {\presp} $N_1$ and $N_2$ are functionally equivalent iff:
\begin{enumerate}
 \item $N_1$ and $N_2$ are cardinality equivalent,
 \item The token values in out-ports in $N_1$ and $N_2$ are the same.
\end{enumerate}
 
 \begin{figure}[htbp]
\centerline{\epsfig{figure=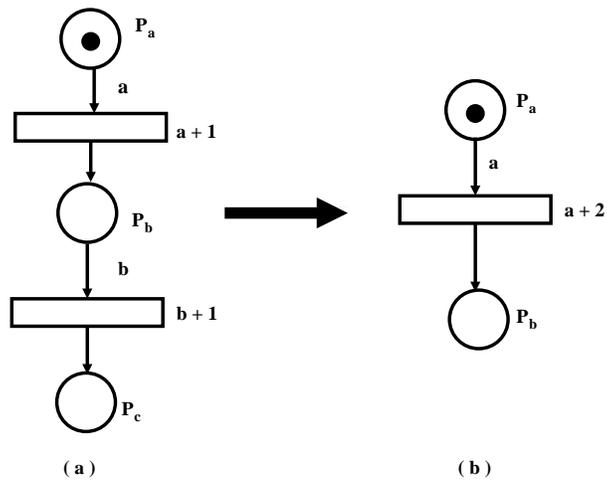,height=63mm}}
\caption{Functional equivalence nets}
\label{fig:t}
\end{figure}
 For example in Figure \ref{fig:t} $N_1$ and $N_2$ are cardinality equivalence. If $P_a$ of $N_1$ and $P_{aa}$ of $N_2$ contain token whose values are 2.
then after execution of $N_1$ and $N_2$ the out-port of $N_1$ and $N_2$ contains token whose values are 5. Hence two nets are totally equivalence.

\subsection{Modeling of a real life example}
\begin{figure}[htbp]
\centerline{\epsfig{figure=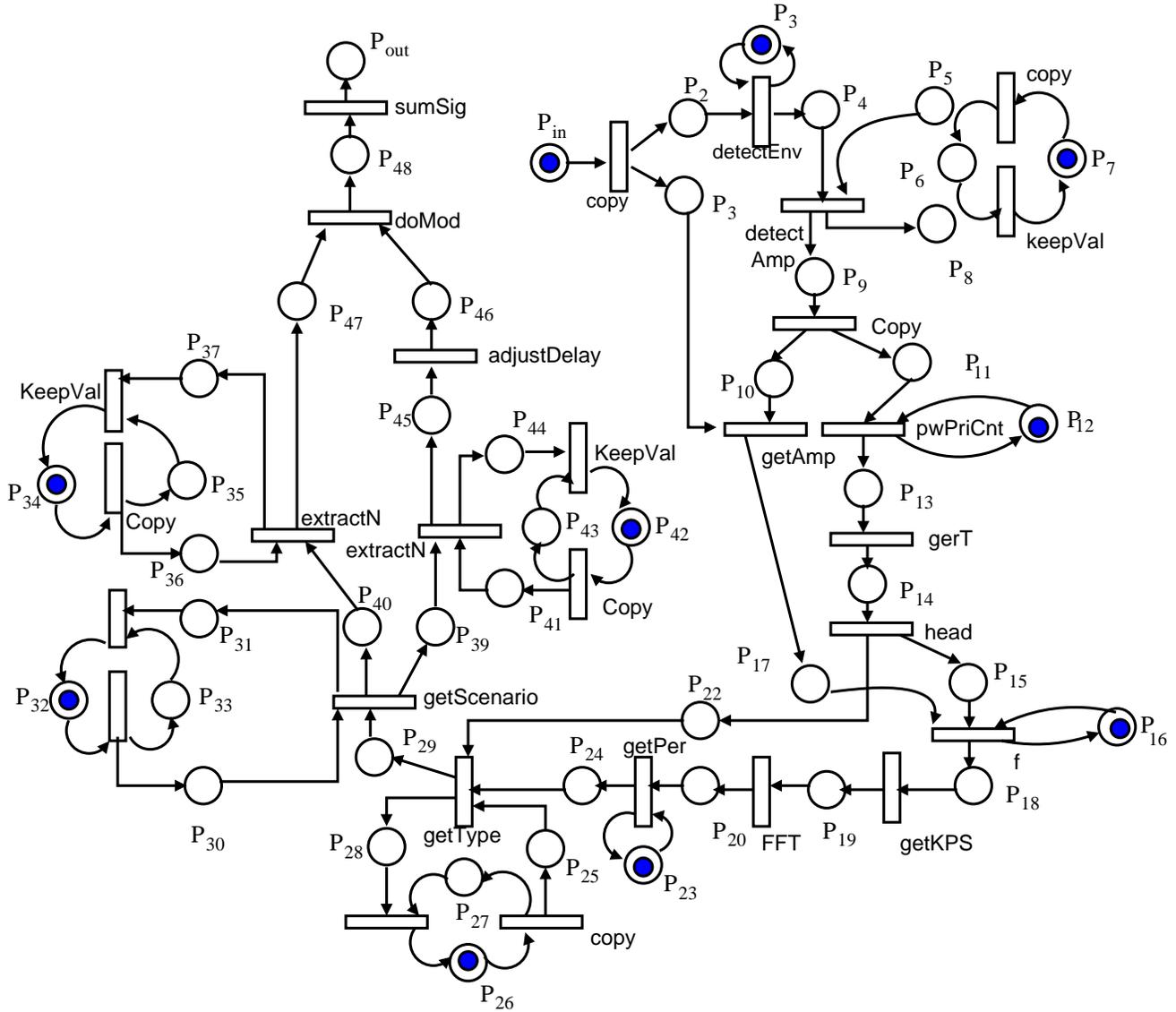,height=150mm}}
\caption{A non pipelined {\presp} model for a jammer}
\label{fig:nonp}
\end{figure}

\begin{figure}[htbp]
\centerline{\epsfig{figure=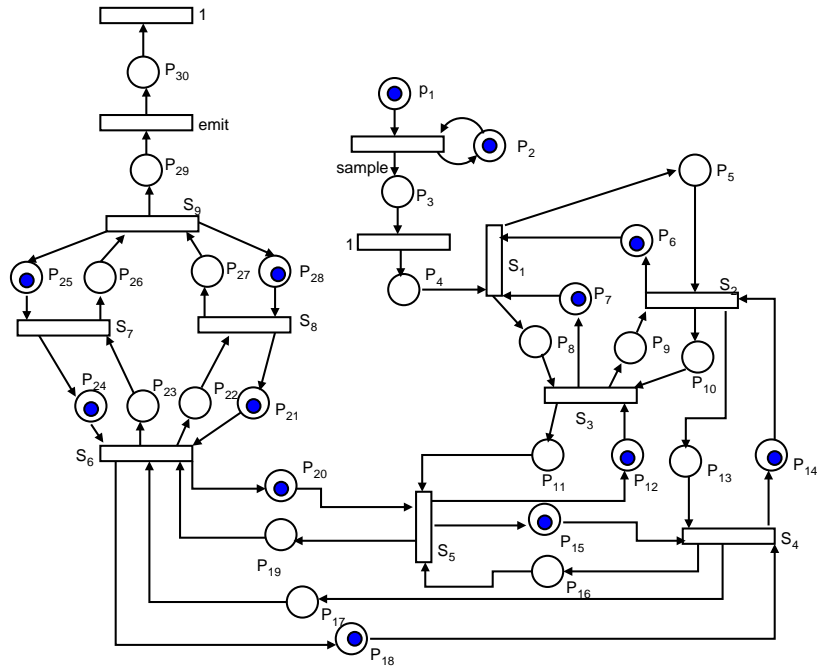,height=88mm}}
\caption{A pipelined {\presp} model for a jammer}
\label{fig:papline}
\end{figure}

Non-pipelined pipelined version of {\presp} model for a jammer is reported \cite{zebo}. Transformation technique from non-pipelined version of {\presp} model 
to pipeline version of {\presp} model also have been reported \cite{zebo}.
Non-pipelined and pipelined version of {\presp} models are shown in  Figure \ref{fig:nonp} and Figure \ref{fig:papline} respectively.

\clearpage
\section{Experimental results}\label{results}

\begin{figure}[htbp]
\centerline{\epsfig{figure=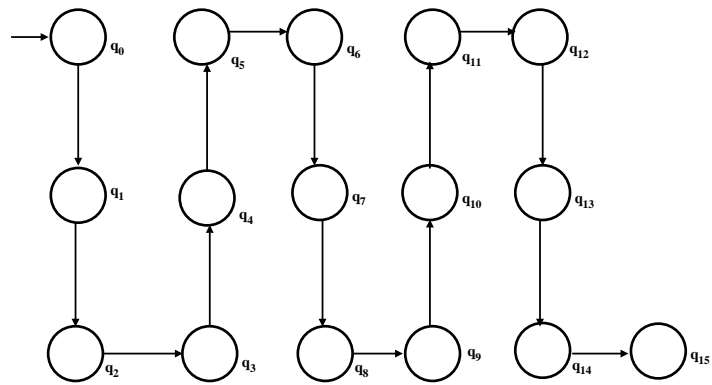,height=50mm}}
\caption{A non pipelined {\fsmd} model for a jammer}
\label{fig:jfsmd}
\end{figure}

\begin{figure}[htbp]
\centerline{\epsfig{figure= 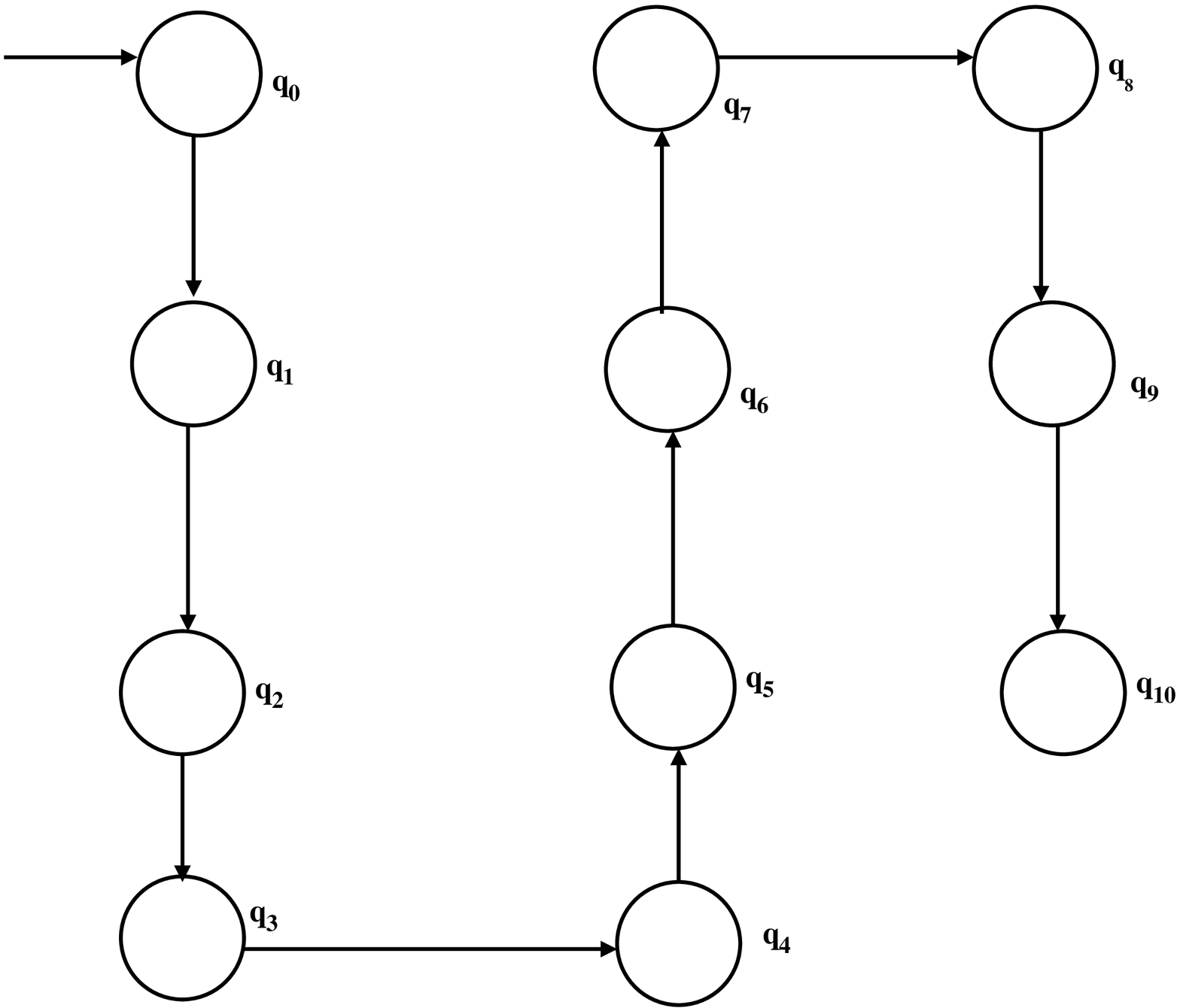,height=50mm}}
\caption{A pipelined {\fsmd} model for a jammer}
\label{fig:pjammer}
\end{figure}

We have reported a translation algorithm from untimed {\presp} model to {\fsmd} model. Hand execution of
this translation algorithm we have get {\fsmd}  model of the jammer from non pipelined {\presp} model. 
The {\fsmd} model is given Figure \ref{fig:jfsmd} and transition function is given in Table \ref{F:comp1}.
\begin{table}[t]
  \begin{center}
    \small\addtolength{\tabcolsep}{-5pt}
\begin{tabular}{|c|c|} \hline 
\emph{State}                           & \emph{Transition function}              \\
               
\hline
$\langle$ $q_0$, $q_1$ $\rangle$         &  in-Copy, Thresold-copy, trigerselect-Copy, opMode-Copy,  modParLib-Copy and delayPerLib-copy   \\ 
 \hline                                              
$\langle$ $q_1$, $q_2$ $\rangle$         &  detectEnv                                                                 \\ 
\hline
$\langle$ $q_2$, $q_3$ $\rangle$         &   detectAmp                                                                                  \\  
\hline
$\langle$ $q_3$, $q_4$ $\rangle$         &  thresold-keepVal, copy                                                                               \\    
\hline
$\langle$ $q_4$, $q_5$ $\rangle$         &  getAmp, pwPricnt                                                                              \\             
\hline
$\langle$ $q_5$, $q_6$ $\rangle$         & getT                           \\ 
\hline
$\langle$ $q_6$, $q_7$ $\rangle$         & head                                                                \\ 
\hline
$\langle$ $q_7$, $q_8$ $\rangle$          & f                                                                                                       \\
\hline 
$\langle$ $q_8$, $q_9$ $\rangle$         & getKPS            \\
\hline
$\langle$ $q_8$, $q_9$ $\rangle$          & FFT                      \\
\hline
$\langle$ $q_8$, $q_9$ $\rangle$          & getPer\\
\hline
$\langle$ $q_9$, $q_{10}$ $\rangle$        & getType \\
\hline 
$\langle$ $q_{10}$, $q_{11}$ $\rangle$       & trigSelect-keepVal, getScenario\\
\hline 
$\langle$ $q_{11}$, $q_{12}$ $\rangle$        &trigSelect-copy, opMode-keepVal, extractN, extractN \\
\hline 
$\langle$ $q_{12}$, $q_{13}$ $\rangle$         & opmode-copy, delayPerLib-keepVal, modPerLib-keepVal, adjustdelay\\
\hline 
$\langle$ $q_{13}$, $q_{14}$ $\rangle$          & delayPerLib-copy, modPerLib-copy, doMod\\
\hline
 $\langle$ $q_{14}$, $q_{15}$ $\rangle$         & sumsig\\
\hline
\end{tabular}
\end{center}
\caption{Transition function for {\fsmd} model obtain from normal {\presp} model of a jammer}
\label{F:comp1}
\end{table}
Similarly, the {\fsmd} generated from the pipelined {\presp} model  is shown in Figure \ref{fig:pjammer} and the 
state transition function given in Table \ref{F:comp3}

\begin{table}[t]
  \begin{center}
    \small

\begin{tabular}{|c|c|} \hline 
\emph{State}                           & \emph{Transition function}              \\
              
\hline
$\langle$ $q_0$, $q_1$ $\rangle$         &  in-Copy $\Diamond$ detectEnv                                                                          \\ 
 \hline                                              
$\langle$ $q_1$, $q_2$ $\rangle$         &  Thresold-copy $\Diamond$ keepVal $\Diamond$ detectAmp                                                                 \\ 
\hline
$\langle$ $q_2$, $q_3$ $\rangle$         &   in-Copy $\Diamond$ getAmp                                                                                   \\  
\hline
$\langle$ $q_3$, $q_4$ $\rangle$         &  pwPriCnt $\Diamond$ getT $\Diamond$ head                                                                                \\    
\hline
$\langle$ $q_4$, $q_5$ $\rangle$         &  f $\Diamond$ getKPS $\Diamond$ FFT $\Diamond$ getPer                                                                              \\             
\hline
$\langle$ $q_5$, $q_6$ $\rangle$         &trigerselect-Copy $\Diamond$ keepVal $\Diamond$ getType $\Diamond$ opMode-Copy $\Diamond$ keepVal $\Diamond$ getScenario                            \\ 
\hline
$\langle$ $q_6$, $q_7$ $\rangle$         &modParLib-Copy $\Diamond$ keepVal $\Diamond$ extractN  and delayParLibCopy $\Diamond$ keepVal$\Diamond$ extranctN $\Diamond$ adjustDelay         \\ 
\hline
$\langle$ $q_7$, $q_8$ $\rangle$         &  doMod $\Diamond$ sumsig                                                                                                       \\
\hline 
$\langle$ $q_8$, $q_9$ $\rangle$         & emit \\
\hline
\end{tabular}
\end{center}
\caption{Transition function for {\fsmd} model obtain from pipelined {\presp} model of a jammer}
\label{F:comp3}
\end{table}
\clearpage

Here the {\fsmd} equivalence checking is very straightforward. Two versions of {\fsmd}s have only one path and the data transformation 
which have been shown in Table \ref{F:comp1} and Table \ref{F:comp3} are same. Hence two {\fsmd} models are equivalent.

\section{Plan of Future work}\label{future-work}
 Carrying out analysis for correctness of technique, complexity analysis, etc. Direct equivalence checking between two {\presp} models
 Generalization of {\fsmd} models to timed {\fsmd} models.
 We will generalize an {\fsmd} model to timed {\fsmd} model which can capture data path as well as timing behaviour and
 Conversion of {\presp} models to timed {\fsmd} models.

\bibliographystyle{ieeetr}
\bibliography{reff}

\end{document}